# Global Shipyard Capacities Limiting the Ramp-Up of Global Hydrogen Transport

Maximilian Stargardt[a, b, *,] David Kress[a], Heidi Heinrichs[a], Jörn-Christian Meyer[c], Jochen Linßen[a], Grit Walther[c], Detlef Stolten[a, b]

[a)] Forschungszentrum Jülich GmbH; Institute of Energy and Climate Research – Techno-economic Systems Analysis (IEK-3), 52425 Jülich, Germany

[b)] RWTH Aachen University, Chair for Fuel Cells. Faculty of Mechanical Engineering, 52072 Aachen, Germany

[c)] RWTH Aachen University, Chair of Operations Management, School of Business and Economics, 52062 Aachen, Germany

* Corresponding author: Maximilian Stargardt,

## Abstract
Decarbonizing the global energy system requires significant expansions of renewable energy technologies. Given that cost-effective renewable sources are not necessarily situated in proximity to the world's largest energy demand centers, the maritime transportation of low-carbon energy carriers, such as renewable-based hydrogen or ammonia, will be needed. However, whether existent shipyards possess the required capacity to provide the necessary global fleet has not yet been answered. Therefore, this study estimates global tanker demand based on projections for global hydrogen demand, while comparing these projections with historic shipyard production. Our findings reveal a potential bottleneck until 2033-2039 if relying on liquefied hydrogen exclusively. This bottleneck could be circumvented by increasing local hydrogen production, utilizing pipelines, or liquefied ammonia as an energy carrier for hydrogen. Furthermore, the regional concentration of shipyard's locations raises concerns about diversification. Increasing demand for container vessels could substantially hinder the scale-up of maritime hydrogen transport.

## Highlights
- 14 suitable shipyards concentrated in East Asia
- Potential hydrogen transport bottleneck until 2033-2039 (scenario-dependent)
- Up to 53 million cubic meter transport capacity lack in 2035 (scenario-dependent)
- Container vessel demand can hinder maritime hydrogen transport
- Liquefied ammonia could be an alternative overcoming the bottleneck

**Abbreviations:**
AiP – Approval in Principle; APS - IEA's Announced Pledges Scenario; CGT – Compensated Gross Tonnage; GT – Gross Tonnage; IEA – International Energy Agency; $LH_2$ – Liquefied hydrogen; $LH_{2eq}$ – Liquefied hydrogen equivalent; $LNH_3$ – Liquefied ammonia; LNG - Liquefied natural Gas; IMO - International Maritime Organization; NZE – IEA's Net Zero Emission Scenario; OECD - Organisation for Economic Co-operation and Development;

# 1 Introduction

The global maritime transport of fossil energy carriers is common nowadays. Over the past few decades, the infrastructure for the maritime transport of coal, natural gas, crude oil and oil products has matured and proven its effectiveness. For example, 13.4% of global natural gas production was traded as liquefied natural gas (LNG) by oceangoing vessels in 2022 [1]. Whereas the share of fossil fuels must be substantially reduced to fight climate change, the global maritime transport of energy carriers is still likely to play an important role in a future greenhouse gas neutral energy system. This is driven by the fact that not all regions can supply themselves with sufficient or cost-effective renewable energies and, hence, will depend on energy imports, which are at least partially being transported over vast distances of several thousand kilometers [2]. As transport via pipelines is limited to an onshore connection or to a specific economically-maximal offshore distance varying from 2,000 km [3] to 3,000 km [4] depending on the specific assumptions with a maximum sea depth of 5 km [5], transport via maritime vessels is a more versatile option. This transport mode also has more flexibility in contracting as illustrated in the natural gas markets [6], which for instance enabled alternative supplies for Europe after the Russian invasion of Ukraine.

In this context, the question arises of whether enough vessels will be available for trading hydrogen-based energy carriers globally to supply the expected rising demand. However, most studies focus on national hydrogen supply, such as those of Bhandari [7] and Karayel et al. [8], European hydrogen supply like Caglayan et al. [9], stop at the exporting harbor [2], or simply assume the sufficient availability of vessels [10–17]. In addition, studies investigate the maritime transport costs of different hydrogen-based energy carriers. Johnston et al. [10] compare the transport cost of liquefied hydrogen ($LH_2$), liquefied ammonia ($LNH_3$), liquefied organic hydrogen carrier (LOHC), and methanol from Australia to the Netherlands. Meca et al. [18] investigate the differences in the transport cost of liquid hydrogen and methanol. Gallardo et al. [11] analyzed renewable based supply chains of $LH_2$ and $LNH_3$, including transport from Chile to Japan. The assumed transport vessels have a capacity of 160,000 m³ for $LH_2$ and 53,000 m³ for $LNH_3$, respectively. Ishimoto et al. [12] examine the production of hydrogen and ammonia in Norway and its transport to Rotterdam in Europe and to Tokyo in Japan. The assumed size of the tankers varies from up to 172,000 m³ for $LH_2$ and up to 85,000 m³ for $LNH_3$. Song et al. [13] examine the least cost transport options for hydrogen and its derivates, such as ammonia or methanol, from China to Japan. They assigned a transport capacity of 160,000 m³ to $LH_2$ vessels and 38,000 m³ to $LNH_3$ ones. Heuser et al. [14] focus on transporting energy as $LH_2$ and consider the entire supply chain from Patagonia to the import terminal in Japan. The tanker size is assumed to be 160,000 m³. In contrast, Fúnez Guerra et al. [19] consider a fixed cost of 50 €/$t_{LNH_3}$ for transporting $LNH_3$ from Chile to Japan. However, all these studies assume sufficient availability of vessels. The size limitations per vessel are summarized in Table *1*.

| Source | LH$_2$ tanker | LNH$_3$ tanker |
|---|---|---|
| Johnston et al [10] | 160,000 m³ | 160,000 m³ |
| Gallardo et al [11] | 160,000 m³ | 53,000 m³ |
| Ishimoto et al [12] | 172,000 m³ | 85,000 m³ |
| Song et al [13] | 160,000 m³ | 38,000 m³ |
| Heuser et al [14] | 160,000 m³ | - |

Table 1: Applied tanker size for liquefied hydrogen and liquefied ammonia in literature

Similarly, potential limitations in maritime transport capacities are also neglected in global energy system studies. Cronin et al. [15] consider ammonia as a maritime transport option in their global framework, but place no constraints on the maritime transport capacity. Nunez-Jimenez et al. [16] allow the maritime transport of LNH$_3$ and LH$_2$, and consider a detailed representation of maritime transport, including loading and unloading times in ports, but apply no ship fleet constraints. Besides considering LNH$_3$ and LH$_2$, a study from the Hydrogen Council and related companies [17] includes the transport of so-called green steel, which is a solid cargo. The results show a growing need for up to 1,115 maritime vessels in 2050, with 330 vessels for transporting green steel, 280 for LH$_2$, 140 for methanol and 80 for liquefied ammonia and 285 vessels for numerous other energy carriers such as carbon dioxide and synthetic kerosene. However, the study does not consider the potential capacity constraints of global shipyards.

The Energy Technology Perspectives 2023 study from the IEA [20] investigates the historic global LNG and LPG tanker output, identifying Japan, South Korea, and China as producing countries for large, liquefied gas tankers and specifying the need for new tankers in its Net Zero Emission (NZE) scenario, while mentioning that shipyard capacity may not suffice until 2030 but in 2050. The IEA quantifies the requirement for newbuild tankers as 20 LH$_2$ tankers and 170 LNH$_3$ tankers by 2030, and 200 vessels for LH$_2$ and 300 for LNH$_3$ by 2050, following their NZE scenario. They assume that only a few shipyards will be capable of providing the necessary manpower and required steel for the very low temperature of liquid gas transport vessels. However, a quantification of the expected delays and the expected shortage in shipyard capacity as well as potential alternatives is missing. Moreover, the IEA does not provide a quantification of the suitable shipyards and is also lacking a potential outlook on further shipyards that might qualify for the construction of suitable tankers. In addition, the assumptions have not been published, and no constraints can be derived from the published findings.

Therefore, this paper aims to quantify and assess the spatially resolved global production potential of shipyards for liquid gas tankers based on historical data. Thus, potential gaps in required maritime transport capacities can be revealed and potential solutions for gaps identified. For this, the transportation capacities for $LH_2$ and $LNH_3$ are investigated. In contrast, further mentioned carrier options, like methanol that is transported in chemical tankers [20], and LOHC that is transported in oil tankers after minor adaptations [4], are not considered in this study.

## 2 Methodology

This study aims to determine the potential global maritime transport capacity which is limited by production capacity of shipyards for large liquefied gas tankers and existing fleets. Figure *1* illustrates the main steps for calculating future maritime hydrogen transport capacity.

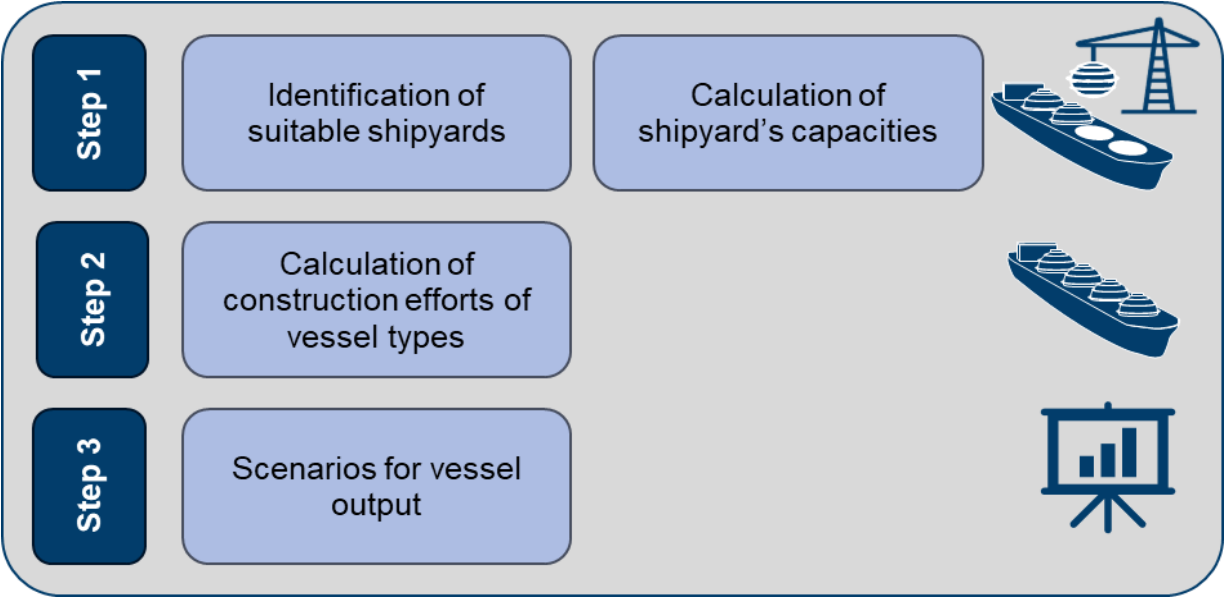

Figure 1: Graphical scheme of the main steps to calculate future maritime transport capacity

First, sufficiently large shipyards are identified as being capable of constructing liquefied gas carriers with predefined dimensions. Following identification of those, their output capacity is derived. Then, the effort to construct LNG, $LH_2$, and $LNH_3$ vessels is determined. Finally, scenarios are generated to examine the potential future output of the shipyards. Each step of the methodology is explained in the following chapter.

### 2.1 Identification of suitable shipyards

Shipyards can theoretically produce different types of vessels. To determine the global shipyard capacities and their dimensions for constructing large liquefied gas tankers, the existing fleet of LNG tankers serves as a baseline. Within the scope of this manuscript, the focus is exclusively on large liquefied gas tankers to ensure economies of scale and suitability for long-distance transportation.

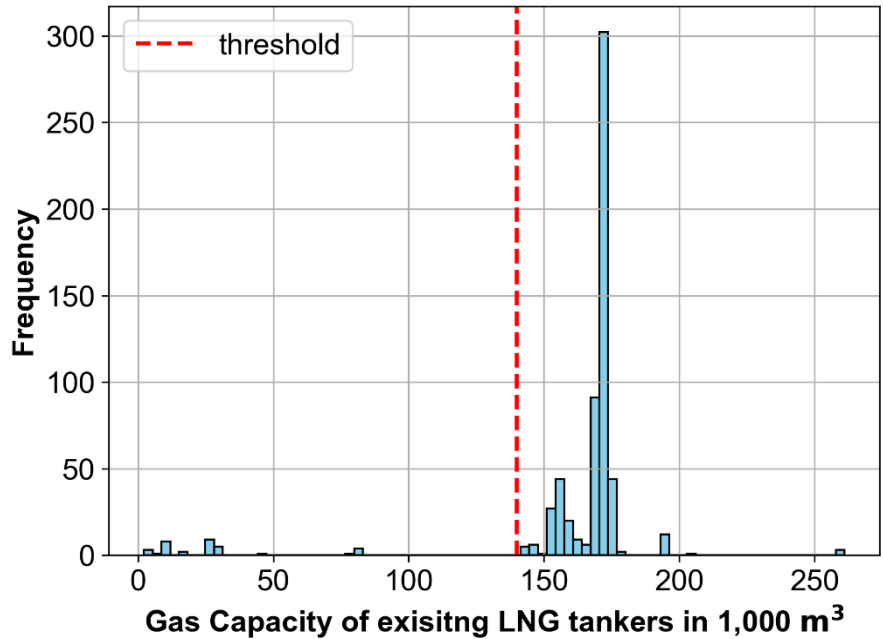

Figure 2. Quantity and capacity of produced LNG tankers since 2010 [21]

All LNG tankers, that have been built since 2010, are considered in this study. Therefore, shipyards that closed due to the Financial Crisis are already excluded. The remaining global LNG fleet can be divided into two groups: LNG vessels with a cargo capacity up to 80,000 m³ and LNG vessels with a minimum cargo capacity of 140,000 m³ [21], as shown in Figure 2. The latter is applied as a lower limit when selecting suitable shipyards within this analysis. Furthermore, the histogram underlines the frequency of cargo capacity classes of constructed LNG tankers which peaks at a cargo capacity of 160,000 m³. If a shipyard was capable of constructing large LNG tankers since 2010, it is identified as a potential construction site for large liquefied gas tankers. LNG tankers are used as a reference because the expertise and infrastructure required from shipyards are expected to be similar for $LH_2$ and $LNH_3$ vessels due to cargo temperature and safety measures for global transportation. To ensure that all identified shipyards are still operating today, shipyards must have built at least one vessel within the last three years (2020-2022), regardless of its type and its size. Three years as a time period are chosen due to the fact that this is the average production time of a defined LNG tanker [21].

## 2.2 Calculation of shipyard capacities for the production of large liquefied gas tankers

A suitable method must be developed, to determine a shipyard's capacity available for vessel construction. Various indicators, such as capacity, deadweight, or displacement, are used to describe vessels sizes [21]. As these indicators are limited upon the vessel type, a comparison of shipyard construction capacity on this basis is not feasible. Another indicator is gross tonnage (GT) which is dimensionless and considers the enclosed spaces of a vessel to calculate the volume and further process that to the stated indicator [22]. However, to compare shipyards and their vessel output, the Organization for Economic Cooperation and Development's (OECD) Compensated Gross Tonnage (CGT) concept (see equation (1)) [23] is employed for the GT of the vessels.

$$CGT = A \times GT^B \qquad (1)$$

Varying workloads required in the construction phase for different vessel types are considered. The parameters A and B are defined for each vessel type based on historical data. Table 2 shows the selected parameters of this approach.

| Vessel type | Parameter A | Parameter B |
| --- | --- | --- |
| LNG tanker | 32 | 0.68 |
| LPG tanker | 62 | 0.57 |
| Oil tanker | 48 | 0.57 |
| Chemical tanker | 84 | 0.55 |

Table 2: Selected vessel types and their parameters in the OECD's CGT approach [23]

By processing the current global fleet and the assignment of the vessels to their corresponding building shipyard, the average annual production capacity of each yard is derived for the period 2015 to 2022. The average construction time for LNG tankers is three years [21]. By considering the above defined period of eight years, a coverage of a full shipbuilding period prior to the COVID-19 pandemic, which disrupted several global supply chains, is ensured. All calculations are based on the global fleet database of S&P Global Market Intelligence [21], which includes all vessels that are more than 100 GT, as defined by International Maritime Organization (IMO) [24]. The available database contains vessels until 2022, limiting the analysis to that year. For the sake of simplicity, and since all tankers use tank systems to contain their cargo, the construction time for LNG tankers of three years is used for all tankers. Thus, starting construction of a vessel in 2024 would allow that vessel to enter service in 2027 at the earliest. In the following parts of the manuscript, it is assumed that the necessary auxiliary industry has sufficient capacity to fulfill the necessary supplies, e.g. for the construction of the required tanks.

### 2.3 Calculating the construction efforts of analyzed vessel types

As previously stated, the CGT concept is employed to assess the construction effort of various vessel types requiring GT data for the vessels. In the context of this manuscript, the CGT must be calculated for a representative LNG tanker as well as for $LH_2$ and $LNH_3$ vessels in order to derive the maximum output of each shipyard per vessel type.

The subsequent section will elaborate on the afore-mentioned vessel types. Furthermore, the typical cargo capacities of the relevant tanker types that are processed within the manuscript are derived. It is important to note that the calculation within this manuscript deals with gas capacity per vessel in terms of transport capacity and not CGT or GT per vessel. To translate this into the required shipyard capacity, a regression analysis for gas cargo capacity and CGT is performed by utilizing the aforementioned global fleet database [21].

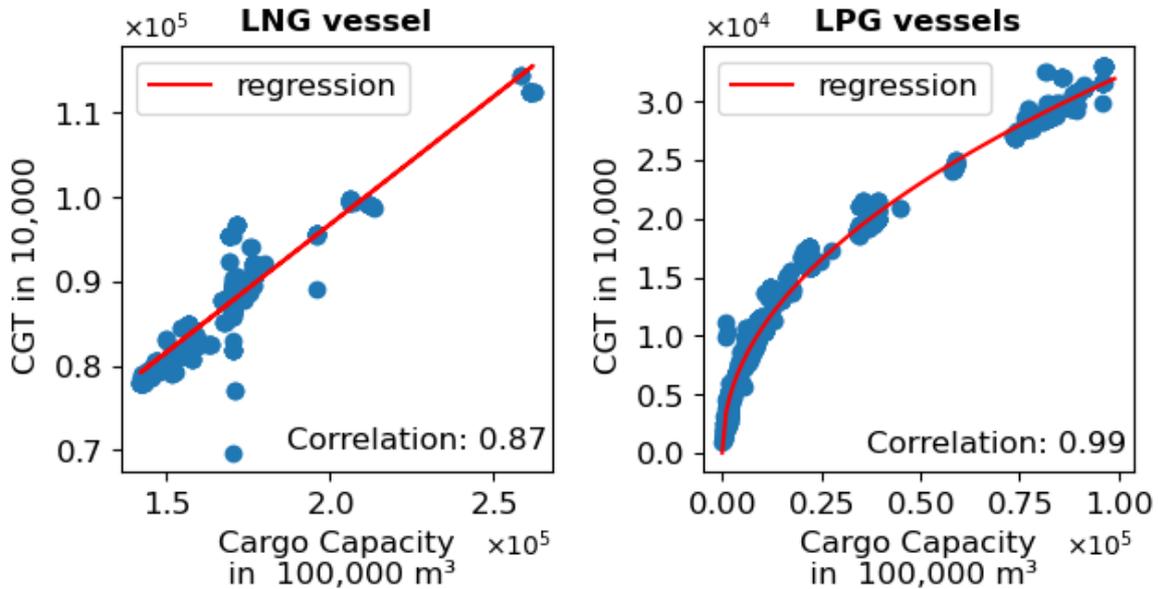

*Figure 3: Correlation between compensated gross tonnage and cargo capacity for LNG vessels (left) and LPG vessels (right)*

**Liquefied natural gas tankers**

The LNG tanker operation plays a significant role in the global natural gas economy, with 660 vessels operating in 2022 [21]. The current average LNG tanker size is around 160,000 m³ of cargo capacity per vessel [6,21]. Due to their better cargo space utilization, compared to other potential tank systems such as self-supporting independent tanks [25,26], the exclusive production of membrane tanks is assumed for new LNG tankers [25,26]. Moreover, cargo membrane tanks are the major tank types in recently produced LNG tankers [6]. A linear regression of LNG tankers with a membrane cargo tank system and a minimum cargo capacity of 140,000 m³ reveals a correlation value of 0.87 regarding the parameters CGT and cargo capacity. This regression leads to equation (2) and is illustrated in *Figure 3*. This equation allows the conversion from a given gas capacity of LNG tankers and $LH_2$ tankers into CGT. Only are considered. A regression following a power function leads to a correlation factor of 0.82 and, therefore, is not considered.

$$CGT = 0.3 \times gas\ capacity + 36{,}087.8 \qquad (2)$$

Since the GT of LNG tankers varies even for similar cargo capacities due to different vessel design concepts, the application of equation (2) allows the derivation of the CGT of LNG tankers using the required gas capacity.

**Liquefied hydrogen tankers**

In 2022, Japanese companies launched the "Suiso Frontier" with a cargo capacity of 1,250 m³ [27] to investigate the transport behavior of $LH_2$ on the high seas. This vessel operates the regular shipments between Australia and Japan. As of 2024, it is the only existing prototype of an $LH_2$ tanker [28]. In addition, Approval in Principle (AiP) has been granted by certification companies such as DNV and Lloyds Register [29] to ensure that the designs comply with current regulations for the construction of $LH_2$ tankers, confirming a loading capacity of 160,000 m³ for $LH_2$ tankers. These AiPs cover designs with membrane and spherical cargo tanks developed by companies in South Korea and Japan [29,30]. In line with these AiPs, a cargo capacity of 160,000 m³ is assumed for $LH_2$ tankers in this manuscript. However, compared to LNG tankers, additional insulation is required for $LH_2$ transport vessels [31]. While a tank insulation thickness of 40 cm is sufficient to handle the LNG cargo with current membrane technology [32,33], a wall thickness of 100 cm is assumed for $LH_2$ tankers to handle the $LH_2$ cargo at -253°C [34]. This increased insulation thickness is considered when calculating the CGT for $LH_2$ tankers. Therefore, an $LH_2$ tanker with a simplified membrane tank system of four cubic membrane tanks of 40,000 m³ capacity each is assumed. Comparing the tanker dimensions and tank sizes, the GT of an $LH_2$ tanker including cargo space and the insulation layer with a total cargo capacity of 160,000 m³ is equivalent to an LNG tanker with a cargo capacity of 43.614 m³ per cargo tank and a total cargo capacity of 174,457 m³ total cargo capacity. Given that $LH_2$ tankers are perceived to be comparable in their construction to LNG tankers [35], the CGT per $LH_2$ tanker is derived using equation (2). In contrast to the calculation of CGT of an LNG tanker with 160,000 m³ gas capacity, the CGT of an $LH_2$ tanker is derived by applying a gas capacity of 174,457 m³ to equation (2).

**Liquefied ammonia tankers**

Due to the physical and chemical similarity of ammonia and liquefied petroleum gas (LPG), ammonia is usually transported in liquid form at -33°C. $LNH_3$ tankers are also suitable for transporting LPG [36]. However, it is important to note that ammonia is a highly toxic substance and must be handled with care. In case of any hazards during maritime transport, there could be a risk of environmental pollution and a drastic impact on surrounding human life and the environment. Currently, $LNH_3$ is transported in large tankers with a transport volume of up to 85,000 m³ [36]. However, $LNH_3$ tankers of larger sizes were ordered in May 2022 with independent cargo tanks and a cargo capacity of 93,000 m³ [37]. No technical justification can be found for the smaller cargo capacity compared to LNG or $LH_2$ vessels, even when cargo density and draught are considered. Crude oil tanker could potentially carry more weight due to the higher density of crude oil compared to $LNH_3$. The most likely reason is therefore economical. For this reason, and to facilitate comparison, a cargo capacity of 160,000 m³ is assumed for LNH$_3$ tankers. This is in line with maximum values in the literature [10,38]. Equation (3) is derived by a regression and includes LPG and LNH$_3$ tankers of the global fleet [21]. The CGT from GT for LNH$_3$ vessels is calculated from GT by applying the parameters for LPG carriers in Table *2* in equation (1), since LPG tankers are commonly transported in independent tanks. The application of equation (3) yields the CGT from a given cargo capacity of related tankers. The regression shows a correlation value of 0.99 and is illustrated in *Figure 3*. A linear regression would solely result in a correlation factor of 0.95.

$$CGT = 126.97 \times gas\ capacity^{0.48} \quad (3)$$

In general, the construction of a tanker designed to carry cryogenic cargo, as it is the case for LNG and LH$_2$ tankers, requires more efforts during the construction phase compared to LNH$_3$ tankers which use a cargo temperature of -33°C.To compare different production portfolios regarding the available shipyard capacity, the unit "liquefied hydrogen equivalents" (LH$_{2eq}$) is introduced. LH$_{2eq}$ is defined as the quantity of energy that is stored in one cubic meter of LH$_2$. It takes into account the different energy densities of LH$_2$ and LNH$_3$, and their energy content. In addition, the hydrogen losses [12] that occur during the reconversion from ammonia to hydrogen are also considered. Thus, an LNH$_3$ tanker with a cargo capacity of 160,000 m³ LNH$_3$ corresponds to a transport capacity of 190,217 m³ LH$_{2eq}$.

## 2.4 Scenarios for future maritime transport capacities of hydrogen

In order to project the future maritime hydrogen transport capacities, the future production portfolio of the identified shipyards must be estimated. For this purpose, scenarios are set up. As there are different types of tankers competing for shipyard capacity, beyond large liquefied gas tankers, a total of six scenarios are considered. The assumptions of the scenarios are summarized in Table *3*. All scenarios assume that the annual available shipyard capacity will remain constant until 2050. Only the identified shipyards can build large tankers for the transport of LNG, LNH$_3$, and LH$_2$. Depending on the applied scenario, the shipyard capacity is first used to meet the production demand for scenario-dependent other vessel types, and then to produce LNH$_3$ and LH$_2$ tankers. The current fleet is also taken into account to meet the requisite transport demand. Therefore, a depreciation period of 25 years for the current fleet is assumed. This is in line with the average vessel lifetime between 20 [10] and 30 years [39] which is reported in the literature.

| Scenario name | LNG transport demand | Further specifications |
|---|---|---|
| *LNG-first* | NZE from IEA | Meet LNG transport demand |
| *Lower hydrogen demand* | APS from IEA | Decreased hydrogen demand |
| *Hydrogen priority* | Neglected | Focus on $LNH_3$ and $LH_2$ tankers |
| *Repurpose shipyards* | NZE from IEA | Extension of shipyards capable of producing liquefied gas tankers |
| *Crude oil inclusion* | NZE from IEA | Available shipyard capacities of crude oil tankers |
| *Container inclusion* | NZE from IEA | Additional capacity demand for container vessel construction |

Table 3: Overview of scenario specifications

Given that shipyards have the autonomy to determine their production portfolio, a solution space is derived. This convex solution space indicates all combinations of $LNH_3$ and $LH_2$ tankers that may be constructed. To assess the potential output of shipyards in relation to the required hydrogen transport capacity demand, it is necessary to derive the latter. As the evolution of a hydrogen market and its shares of maritime transport is uncertain, the same share as for maritime-transported LNG to total demand of natural gas is applied to global projected hydrogen demand [40,41] to derive the share of maritime transport demand for hydrogen. The data on global hydrogen demand is only available for 2030 and 2050. Therefore, the demand is linearized and compared to the potential shipyard output with the required maritime transport capacity for hydrogen from 2030 to 2050. This allows for the identification of potential bottlenecks in transport capacities.

**LNG-first scenario**
The *LNG-first scenario* considers the projected LNG and hydrogen demand within the Net Zero Emission (NZE) scenario of the IEA [40,41]. The NZE scenario represents a transformation pathway for global energy supply with the objective of limiting global warming to the 1.5 °C goal by 2050. Consequently, the global LNG demand is projected to decline from 486 billion m³ LNG in 2025 to 153 billion m³ annually by 2050 [40,41]. The aforementioned NZE scenario encompasses global hydrogen demand of 210 Mt hydrogen and 530 Mt hydrogen in 2030 and 2050, respectively. The resulting maritime transport capacity demand is depicted in Appendix B1.

**Lower hydrogen demand scenario**
In contrast, the *lower hydrogen demand scenario* is based on the Announced Pledges scenario (APS) from the IEA. This considers publicly announced governmental pledges in terms of energy transition and anticipates a reduction in LNG demand from 510 billion m³ of annual LNG demand in 2025 to 324 billion m³ of annual LNG demand in 2050. The scenario represents an energy system transition that does not align with the 1.5°C goal in terms of global warming Both LNG demands are shown in Figure *4*.

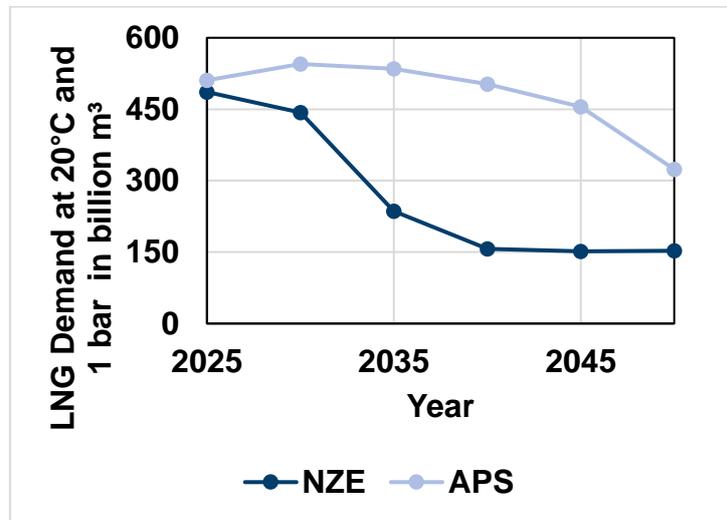

Figure 4. Global LNG demand from 2025 until 2050 according to IEA's NZE and APS scenarios [40]

In contrast to the aforementioned *LNG-first scenario*, the global hydrogen demand in the APS is estimated to reach 130 Mt and 250 Mt in 2030 and 2050, respectively. The derived maritime transport capacity demand is depicted in Appendix B1.

**Hydrogen priority scenario**
*The hydrogen priority scenario* is deficient in its consideration of the required LNG transport capacity and instead prioritizes the production of $LNH_3$ and $LH_2$ tankers.

**Repurpose shipyards scenario**
The repurpose shipyards scenario encompasses all shipyards that have previously constructed LNG tankers, continue to operate in the present day, and have the capacity to produce vessels of defined sizes.

**Crude oil inclusion scenario**
The identified shipyards do not solely produce liquefied gas tankers; rather, they also allocate a portion of their available shipyard capacity to the construction of crude oil tankers and container vessels [21]. Therefore, the global crude oil demand of the IEA's NZE scenario is translated into required crude oil transport capacity in a manner analogous to the approach used for hydrogen and LNG demand. No further crude oil tankers are required until 2035. From 2035 to 2050, the necessity for the production of new crude oil tankers is derived. It is assumed that the total identified shipyards will construct the same proportion of global crude oil transport capacity as they did in the past, within their production portfolio. The remaining shipyard capacity will then then be utilized to produce LH$_2$ and LNH$_3$ tankers in the *crude oil inclusion scenario*.

**Container inclusion scenario**

In addition to tankers, the identified shipyards produce container vessels as well. To estimate the requirement for future global container vessel demand, the global GDP is applied, as it shows a significant correlation to global maritime container transport capacity with a correlation factor of 0.97. The current container vessel fleet's transport capacity is derived [21,42], and future transport demand is derived through the application of a GDP growth of 2.4% per year, following the baseline scenario of the OECD [43]. The additional required shipyard capacity results in a reduction in the shipyard capacity that was intended to produce LH$_2$ and LNH$_3$ tankers in the *container inclusion scenario*.

## 3 Results and Discussion

The results provide insights into the potential for constructing large LH$_2$, LNH$_3$, and LNG tankers. First, the spatial distribution of identified shipyards capable of constructing such vessels is shown. Moreover, their corresponding theoretical output capacity is presented, providing a foundation for the scenarios. The results of the scenarios are illustrated afterward.

### 3.1 Global distribution of shipyards and production capacities

Since 2015, a total of fifteen shipyards globally have produced large LNG tankers that meet the definition outlined in this manuscript. Fourteen of those shipyards still exist today. South Korea has five sites capable of producing LNG tankers, China and Japan have four each and Russia has one. Figure 5 represents the spatial distribution of the shipyards and their average annual share of global tanker output considering the global CGT. The five identified South Korean shipyards contribute to 44.2% of global tanker production. The identified Chinese and Japanese shipyards constructed 4.4% and 4.3% of global tanker output, respectively. he one identified Russian shipyard contributes 0.7% of global tanker production. Therefore, South Korea is identified as the main country constructing tankers. The remaining 46.4% of global tanker production is attributed to other shipyards that do not meet the selection criteria. This production encompasses all sizes and types of tankers.

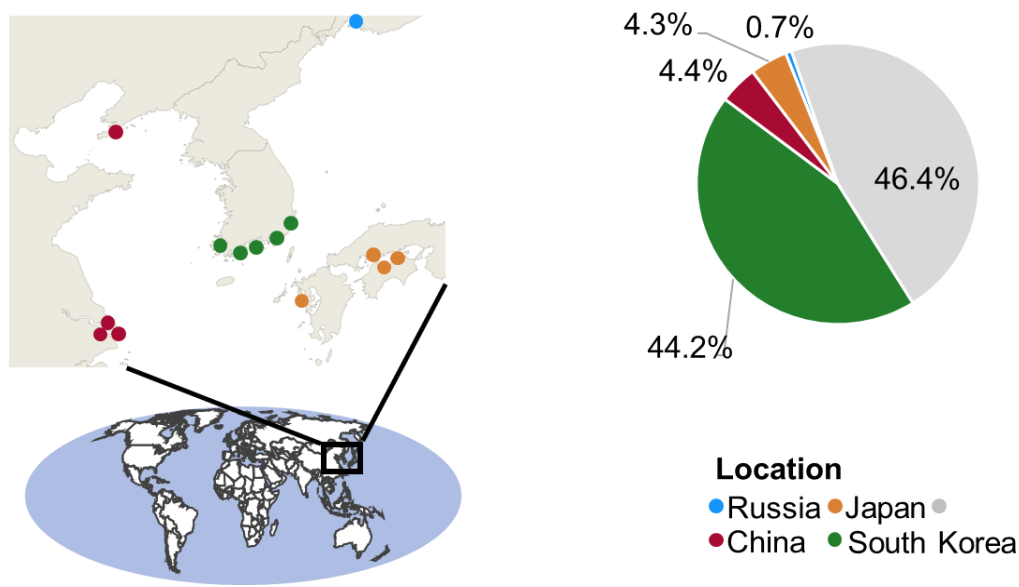

Figure 5: Spatial distribution of global shipyards producing large tankers and their share in overall global tanker production regarding CGT

A quantitative analysis of the output of the fourteen identified shipyards reveals that the tanker production portfolio can be differentiated into four distinct types of tankers. As illustrated in Table 4, the combined production of LNG tankers by all identified shipyards is estimated to be 3.3 million CGT per year between 2015 and 2022. In addition, crude oil and LPG tankers are produced with 2.1 million CGT and 0.7 million CGT, respectively. As can be seen, tankers for chemical and oil products do not play a significant role in the tanker production portfolio.

| Tanker type | Compensated gross tonnages [-] |
|---|---|
| LNG | 3,330,995 |
| Crude oil | 2,120,001 |
| LPG | 662,419 |
| Chemical/Oil products | 156,793 |
| Oil products | 14,1712 |

Table 4: Identified shipyard's annual average tanker production between 2015 and 2022

Figure 6 illustrates the country-wise potential output of the identified shipyards, which highlights the dominance of South Korea in the construction of large-sized LNG tankers. On average, 2.8 million CGT of LNG tankers were constructed in South Korea alone annually, representing 84% of the global average yearly large-sized LNG tanker output. Although the production of crude oil and LPG tankers is relatively modest, their share of production capacity is considerably higher than in the other three countries. China and Japan also have the capacity to construct large-sized LNG and LPG tankers, whereas Russia's contribution to LNG tanker production is minimal.

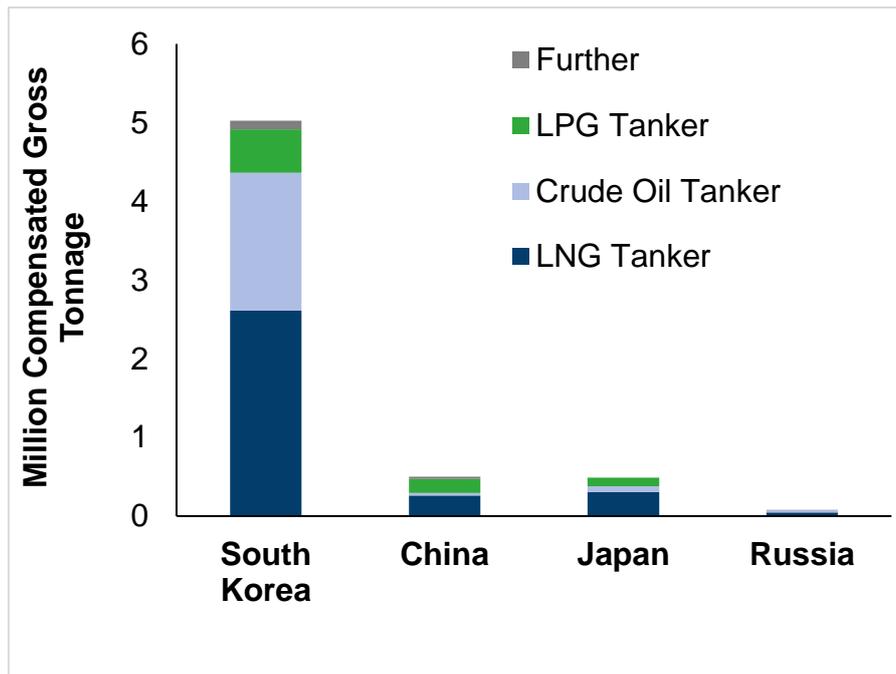

Figure 6: Average annual tanker output of the identified large-sized shipyards between 2015 and 2022

While other regions of the world had previously constructed large LNG tankers before 2010, the shipyards in those areas have either ceased operation or shifted their focus to the construction of other vessel types, such as cruise ships. For instance, Europe has shipyards that have produced LNG tankers. In total, those shipyards constructed 18 LNG tankers between 1994 and 2007, but none have been constructed since then.

## 3.2 Ramp-up of global maritime transport capacity

The results of the scenarios presented in Chapter 2.4 are explained in detail below and illustrated in Figure 8 and Figure 9. The LNG-first scenario is discussed in depth to facilitate comprehension of the subsequent results.

**LNG-first scenario**

As the *LNG-first scenario* is regarded as a benchmark which subsequent discussion may be measured, it is first subjected to analysis. By combining the available annual shipyard capacity for LPG and LNG tankers, the available annual shipyard capacity for the *LNG-first scenario* results in a total shipbuilding capacity of 4.0 million CGT per year.

| Tanker type | CGT per tanker | Total cargo capacity | Maximum number of tankers |
|---|---|---|---|
| LNG tanker | 84,087 | 7.52 million m³ | 47 |
| LH$_2$ tanker | 88,425 | 7.20 million m³ | 45 |
| LNH$_3$ tanker | 39,965 | 15.84 million m³ | 99 |

Table 5: Potential annual production and quantity of transported cargo per tanker type

Table 5 presents the distinct CGTs for each tanker type, as calculated using equations (2) and (3). It also illustrates the projected global annual shipyard output potential of the identified shipyards in the event of a single tanker type being constructed. The thicker insulation of LH$_2$ tankers, which results in a higher CGT, leads to a reduction in the total cargo capacity compared to LNG tankers. However, the potential number of LNG or LH$_2$ tankers is quite similar, whereas the total cargo capacity of LNH$_3$ tankers is much higher. This is due to the CGT covering less than half of LNG or LH$_2$ tankers.

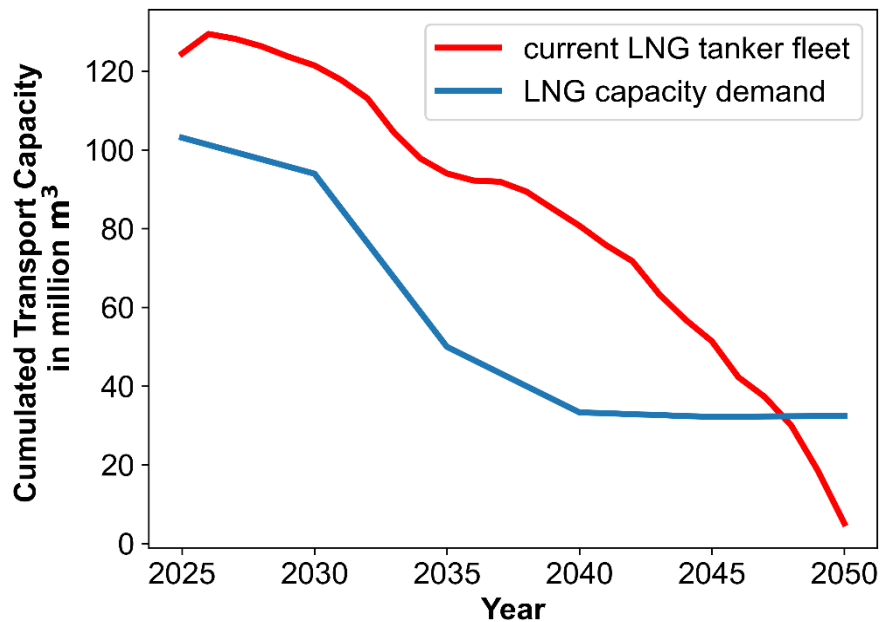

Figure 7: Development of the transport capacity of the current LNG tanker fleet and the required LNG transport capacity demand calculated from the IEA's NZE scenario through 2050

Analyzing the global fleet and global LNG transport capacity derived from the IEA's NZE scenario reveals that no further LNG vessel must be constructed until 2045 (see Figure 7). This is due to the fact that the depreciation of the current LNG fleet requires additional LNG tanker construction beginning in 2047. In total, 171 tankers must be added to the global LNG fleet.

In accordance with the *LNG-first scenario,* the global shipyard capacities are sufficient to meet the hydrogen transport capacity in 2050, regardless of the vessel type transporting the hydrogen. This is demonstrated in Figure 8.

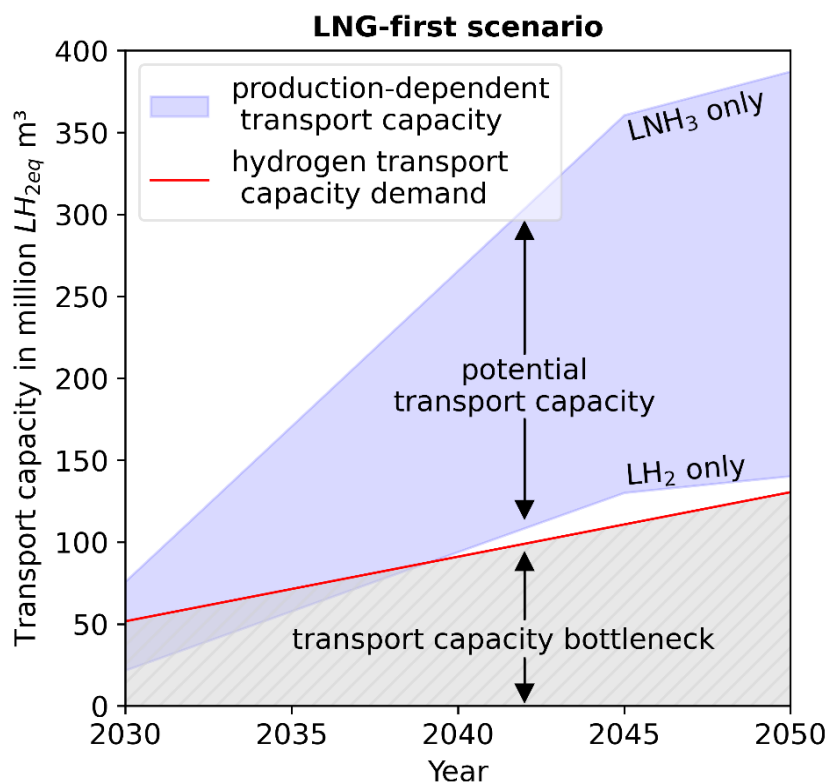

Figure 8: Ramp up of maritime transport capacities for hydrogen in the *LNG-first scenario*

In general, the lower bound of the solution space, which is referred to as the production-dependent transport capacity, is determined by the possible production portfolio that exclusively comprises $LH_2$ tankers. Conversely, the upper bound is determined by a production portfolio that includes only $LNH_3$ tankers. As previously outlined in section 2.4, the requisite maritime hydrogen transport demand is derived from the current share of LNG transport in natural gas markets for the specified values in 2030 and 2050. The derived values for the maritime hydrogen transport capacity demand are included in Appendix B.1 and are depicted as a red line in the figure. Maritime transport production portfolios that do not exceed the global hydrogen capacity demand result in a capacity bottleneck, as by the gray area in the figure. Consequently, the exclusive use of $LH_2$ tankers leads to a shortage of transport capacity from 2030 that will only be resolved around the year 2039. To fulfill this unmet demand, hydrogen must be transported in a manner that includes $LNH_3$, other transport modes, or an increasing local hydrogen production adjacent to the hydrogen demand is required.

Table *6* presents the underlying numbers of the *LNG-first scenario.* In general, the calculation is performed in five -year steps, with the five years preceding the aforementioned year being cumulated. Thus, the depicted value of 19.97 million CGT available shipyard capacity for 2035 refers to the period between 2031 and 2035. As outlined, the global demand in the IEA NZE scenario is not expected to require any newly built LNG tankers before 2046. Consequently, the derived shipyard capacity of 3.99 million CGT per year leads to a cumulated five-year capacity of 19.97 million CGT. It is projected that $LH_2$ tankers will be commercially available in 2028, whereas the $LNH_3$ tankers could already be available in 2027, due to a three-year construction period starting from today. Therefore, the available shipyard capacity is differentiated for both tanker types and decreases due to the required LNG transport capacity demand in 2050.

| Year | 2030 | 2035 | 2040 | 2045 | 2050 |
|---|---|---|---|---|---|
| Available shipyard capacity $LNH_2$ tanker [million CGT] | 11.98 | 19.97 | 19.97 | 19.97 | 5.605 |
| Available shipyard capacity $LNH_3$ tanker [million CGT] | 15.97 | 19.97 | 19.97 | 19.97 | 5.605 |
| Maximum production of $LH_2$ tankers capacity [million $LH_{2eq}$ m³] | 21.68 | 36.13 | 36.13 | 36.13 | 10.14 |
| Maximum production of $LNH_3$ tankers capacity [million $LH_{2eq}$ m³] | 75.93 | 94.91 | 94.91 | 94.91 | 26.65 |
| Maximum number of newbuilt $LH_2$ tankers | 136 | 226 | 226 | 226 | 64 |
| Maximum number of newbuilt $LNH_3$ tankers | 400 | 500 | 500 | 500 | 141 |

Table 6: Available shipyard capacities and tanker production capacities in the *LNG-first scenario*

The outcomes of the further scenarios presented in Chapter 2.4 are elucidated in depth below, accompanied by a visual representation in Figure 9.

**Lower hydrogen demand scenario**

A comparison of LNG and hydrogen demands in the lower hydrogen demand scenario, in light of the announced national policies as summarized in the IEA's APS scenario, reveals a lower required transport capacity by 2050 compared to the *LNG-first scenario*. This is illustrated in Figure 9a. It is important to note that the displayed capacity demand is only valid for the *lower hydrogen demand scenario* and not for other investigated scenarios that have a higher hydrogen transport demand. Nevertheless, a capacity deficit persists in the context of global vessel production that is exclusively pertaining to $LH_2$ tankers until 2035. This is attributed to an increased LNG transport capacity demand in comparison to the *LNG-first scenario*. The reduction in the available shipyard capacity for the construction of $LH_2$ and $LNH_3$ tankers represents a further contributing factor. For a detailed analysis of these figures, please refer to Appendix A1.

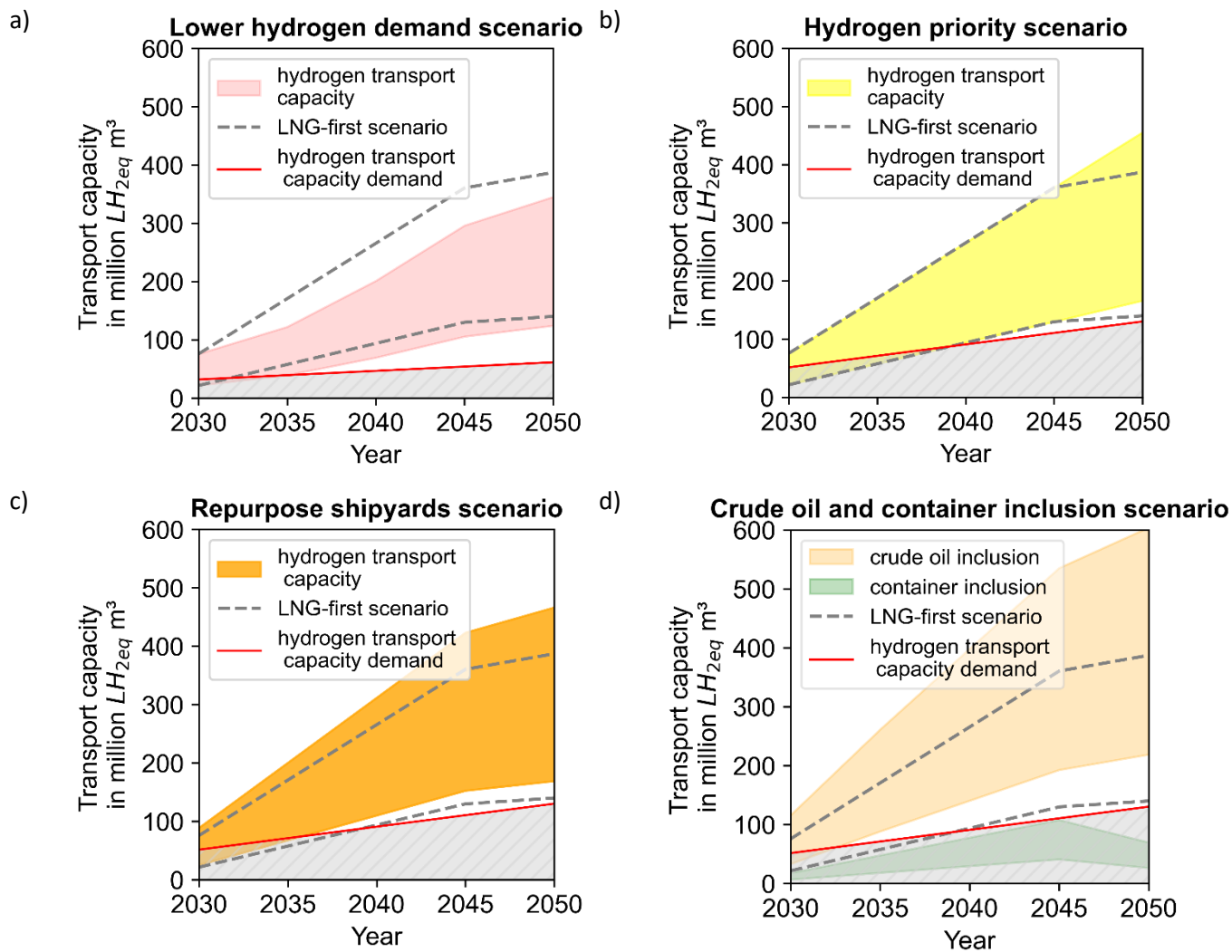

Figure 9: Development of global hydrogen transport capacity for a) lower hydrogen demand scenario, b) hydrogen priority scenario, c) repurpose shipyards scenario, and d) crude oil inclusion scenario and container inclusion scenario.

**Hydrogen priority scenario**

The results of the hydrogen priority scenario are presented in Figure 9b. Neglecting the global LNG transport capacity demand does not alter the potential capacity shortage at the beginning of the 2030s. Nevertheless, as in the *LNG-first scenario*, there is also no capacity shortfall while producing only LH$_2$ and LNH$_3$ tankers from 2039 onward. In comparison to the *LNG-first scenario*, the results of the *hydrogen priority scenario* remain consistent until 2046. However, from that point onwards, the construction of new LNG tankers would be necessary to fulfill the remaining LNG transport capacity demand. Given the surplus of capacity within shipyards from 2045 onwards and the ability to easily meet the LNG tanker demand, this is a viable option. In conclusion, the LNG demand and related LNG transport demand will not have a further impact on the hydrogen transport capacity demands under the assumptions made in this study. The derived parameters for this scenario can be found in Appendix A2.

**Repurpose shipyards scenario**

In the *repurpose shipyards scenario,* repurposing shipyards that have previously produced LNG tankers and can still construct large vessels, does not alter the findings of the *LNG-first scenario*. The results are depicted in Figure 9c. In total, only four additional yards in Asia (two in China, one in Japan, and one in South Korea) are identified that match the requirements. These shipyards represent an additional shipyard capacity of 0.7 million CGT or 3.5 million CGT over five years compared to the *LNG-first scenario*. The detailed derived parameters are shown in Appendix A3. Consequently, the identified shipyards provide a greater capacity that can be applied for $LH_2$ or $LNH_3$ tanker production additionally. In contrast to the *LNG-first scenario,* a global transport infrastructure relying entirely on $LH_2$ tankers would already be able to meet the required transport demand in 2036. However, the capacity shortage before 2036 is not resolved by the inclusion of repurposed yards.

**Crude oil inclusion scenario**

In the event of the *crude oil inclusion scenario*, a reduction in the demand for crude oil tankers will result in an increase in the available shipyard capacities for hydrogen transport. Until 2035, no further crude oil tankers must be constructed, and the maximum yard capacities will be available for the production of $LH_2$ or $LNH_3$ takers. This accounts for an additional 2.1 million CGT shipyard capacity per year, or a 10.6 million CGT shipyard capacity per accumulated five-years capacity. Beginning in 2036, 2.2 million CGT must be applied every five years for the production of crude oil tankers. As previously described in the *LNG-first scenario*, the LNG transport capacity demand must be fulfilled by 2046, further reducing the quantity of available shipyard capacity. The exact numbers are presented in Appendix A4. A comparison of the scenario results with those of the *LNG-first scenario* reveals a significant increase in the constructed maritime hydrogen transport capacity, resulting in an excess capacity by 2033 at the latest. However, as illustrated in Figure 9d, a production portfolio comprising solely of $LH_2$ tankers will still result in a transport capacity shortage by 2033.

**Container inclusion scenario**

In light of the assumptions regarding the global container vessel demand, it is evident that the average global container vessel output must undergo a significant increase in order to align with the requisite global transport demand. It is postulated that the identified shipyards will increase their container vessel production in a uniform manner, corresponding to the required increase in global container vessel production. The maritime hydrogen transport capacity in the container scenario is depicted in Figure 9d. According to the aforementioned findings, the rising demand for container transport capacity, as described in Section 2.4, could be a crucial factor in the competition for available shipyard capacities. The container transport demand decreases the available shipyard capacity by 2.73 million CGT annually in comparison to the *LNG-first scenario* (see Appendix A5). If container vessels were prioritized, the transport capacity for hydrogen would never satisfy demand until 2050, leading to a shortage of maritime transport capacity. Moreover, between 2046 and 2050, there would be no shipyard capacity to produce $LH_2$ or $LNH_3$ tankers due to the necessity of replacing the depreciated container vessels and the further increase of container transport demand. In this scenario, the available maritime hydrogen transport capacity would be significantly reduced by the container demand and the competition regarding scarce shipyard capacities, which could potentially occur.

### 3.3 Consequences of maritime transport limitations

The results of the scenarios indicate that the demand for hydrogen transport capacity cannot be met solely by focusing on $LH_2$ tanker production. To prevent a capacity shortage, the application of hydrogen pipelines or ammonia as an energy carrier may offer a solution. Furthermore, domestic hydrogen production could contribute to meeting the demand. Table 7 presents the minimum number of $LNH_3$ tankers required to meet the hydrogen transport capacity demand in 2030 and 2035, as well as the resulting gap between hydrogen transport capacity and the demand for each scenario.

|  | **Minimum number of $LNH_3$ tankers** | | **Hydrogen transport capacity gap in million $m^3$** | |
| --- | --- | --- | --- | --- |
| **Year** | **2030** | **2035** | **2030** | **2035** |
| *LNG-first scenario* | 255 | 117 | 30.02 | 13.58 |
| *Lower hydrogen demand scenario* | 87 | 1 | 10.33 | 0.05 |
| *Hydrogen priority scenario* | 255 | 117 | 30.02 | 13.58 |
| *Repurpose shipyards scenario* | 225 | 31 | 26.25 | 3.51 |
| *Crude oil inclusion scenario* | 160 | 0 | 18.51 | - |
| *Container inclusion scenario* | - | - | 44.85 | 53.11 |

Table 7: Minimum number of $LNH_3$ tankers for meeting the required hydrogen transport capacity demand and the resulting hydrogen transport capacity demand of only $LH_2$ tankers produced per scenario

The *LNG-first scenario* and the *hydrogen priority scenario* necessitate the highest number of $LNH_3$ tankers of all scenarios to meet the required hydrogen transport capacity demand. By 2030, 355 $LNH_3$ tankers are required, and by 2035, 117 $LNH_3$ tankers are required. In contrast, the *crude oil scenario* does not require any $LNH_3$ tankers by 2035 because the available $LH_2$ tankers can fulfill the required demand. In the *container inclusion scenario*, insufficient quantity of transport capacity leads to capacity gaps of 44.9 and 53.1 million $m^3$, respectively. Furthermore, the *lower* *hydrogen demand scenario* will fail to meet the transport capacity requirement by 0.1 million m³ in 2035.

Figure 10 depicts the prospective global maritime transport capacities for the examined scenarios, illustrating the scenario outcomes for 2030 and the related capacity demand. The upper limit of each scenario's solution frame is determined by a shipyard production portfolio that is solely based on $LNH_3$ tankers, whereas the lower limit is determined by a portfolio that is solely based on $LH_2$ tankers. As the *lower hydrogen demand scenario* pertains to other LNG demands, it is not directly comparable to the other scenarios and, thus, excluded from Figure *10*.

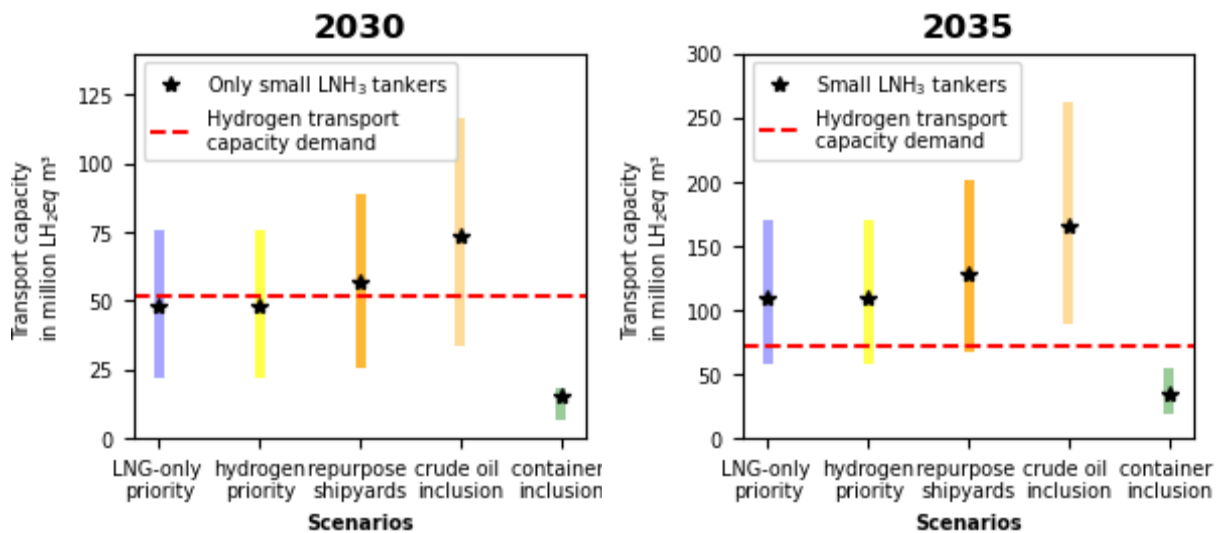

Figure 10: Comparison of transport capacities for the analyzed scenarios in 2030 and 2035 and the impact of small $LNH_3$ tanker

It is evident that the hydrogen transport capacities in 2030 and, therefore, the fulfillment of capacity demand depend on the selected production portfolios. While the availability of shipyard capacity for the construction of dedicated crude oil tankers and the potential for additional yards to build large liquefied gas tankers may increase transport capacity, the utilization of $LNH_3$ for maritime hydrogen transport and, consequently, the construction of additional $LNH_3$ tankers represents a viable alternative to increasing local production to meet anticipated hydrogen demand. In the event that maritime hydrogen transportation is limited to $LH_2$ tankers, the LNG-first scenario and the crude oil inclusion scenario result in a transport capacity gap of between 30.2 million and 18.5 million $LH_2$ m³. This equates to the construction of 189 and 116 $LH_2$ tankers, respectively, to bridge the gap. As the identified shipyards lack the capacity to expand their production, either other shipyard must commence production of the required $LH_2$ vessels or new shipyards must be constructed, thereby increasing the global shipyard capacity.

LNH$_3$ tankers with a cargo capacity of 93,000 m³ are under construction. As mentioned above. However, this research's methodology assumes an LNH$_3$ tanker cargo capacity of 160,000 m³. The use of LNH$_3$ tanker with the cargo capacity of 93,000 m³ cargo capacity affects the solution space in each scenario. According to the power function in equation (2), a smaller cargo capacity of each LNH$_3$ tanker decreases the total cargo capacity of the shipyards. The solution space of each scenario becomes smaller. For this, Figure 10 also shows the solution frames for smaller LNH$_3$ tankers in 2030 and 2035. In 2030, even an exclusive focus on small LNH$_3$ tankers would lead to a maritime hydrogen transportation capacity bottleneck in the *LNG-only priority* and the *hydrogen priority scenarios*. From 2035 onwards, small LNH$_3$ tankers still reduce the solution space, but are still sufficient to meet the demand for hydrogen transport capacity.

The required quantity of maritime container transport capacity may potentially increase the scarcity of available shipyard capacities, thereby intensifying the bottleneck in maritime hydrogen transport capacities. The projected increase in demand for container transport capacity, and thus the demand for additional vessels, will not permit the hydrogen transport capacity to meet demand, regardless of the selected production portfolio between LNH$_3$ and LH$_2$ tankers.

Assuming an average vessel construction time of three years, as described in Section 2.2, a new shipyard must be operational before 2028. In this case, the respective ships would contribute to the resolution of the bottleneck in the early 2030s. In addition, shipyards that have not yet been identified as suitable, could consider expanding their production of crude oil tankers and container vessels in order to offer more shipyard capacities for the production of LH$_2$ and LNH$_3$ tankers. In the event that the dimensions of the shipyard dimensions are insufficient, an extension of the shipyard is required in order to accommodate the potential vessel's depth, overall length, and beam.

The identified shipyards are exclusively located in East Asia. From the perspective of energy supply security, it may be desirable to diversify tanker production. Even in the *repurpose shipyards scenario*, the four additional shipyards that are considered, which are located in East Asia as well. No further shipyards are deemed capable of constructing ships of the required dimensions, as no other shipyard has yet constructed an LNG tanker, regardless of its size.

The incorporation of existing shipyards with the capacity to construct vessels of the requisite dimensions in terms of the vessel size would expand the number of potential yards by 52. However, 45 of these yards would still be in Asia, five in Europe, and one each in the United States and in Brazil. Nevertheless, it is uncertain whether those shipyards will be able to construct liquefied gas tankers in the future, given that the complex construction processes associated with cryogenic cargo handling and the required safety measures. In order to address the described bottleneck in maritime transport capacity, the integration of this process must be completed within the next three years. Moreover, in terms of the security of energy supply, governments from countries outside the East Asian shipbuilding countries should consider encouraging domestic or other overseas industries to undertake liquefied gas tanker production. This would ensure a certain level of diversification in the future, even when neglecting the described bottleneck.

Moreover, a global trade economy could theoretically rely on smaller liquefied gas tankers that are not considered in this research. A decreasing tanker volume would lead to a smaller economy of scale regarding the cost of transport, which is not further examined in this paper.

Furthermore, it is important to note that the conducted analysis relies on a transformation path outlined by the IEA. This implies that the maritime transport capacities are strongly dependent on national energy systems and vice versa. For instance, the IEA assumes a mix of $LNH_3$, and $LH_2$ transport chains combined with the partial conversion of ammonia into hydrogen or power in the destination countries. In contrast, the International Renewable Energy Agency assumes a global ammonia transport chain with no reconversion to hydrogen in destination countries. In this case, national energy systems must be prepared to process ammonia as an energy carrier. However, if combustion processes utilize ammonia, the potential increase in $NO_x$ emissions in exhaust gas streams may be an issue that must be addressed [44].

## 4 Conclusions

The primary objective of the conducted research is to project the available shipyard capacity for the future production of $LH_2$ and $LNH_3$ tankers, considering various factors, including the global LNG demand, the availability of shipyards, and the competition among other vessel types in suitable shipyards. To achieve this objective, this study analyses six scenarios and their impact on maritime hydrogen transport capacity. In order to properly analyze the shipyard capacity, it is first necessary to determine suitable shipyards as a starting point.

The findings of this paper indicate a discrepancy between the projected capacity for the maritime hydrogen transport under current conditions and the projected demand for such transport capacities. The period until the late 2030s is characterized by maritime hydrogen transport bottlenecks. Depending on the chosen scenario, a maritime hydrogen transport gap of up to 53 million m³ could occur when solely focusing on the production of $LH_2$ tankers. To avoid this, a scale-up of $LNH_3$ transport chains is required to fulfill the maritime transport demand under the assumptions of this study. The expertise for constructing large liquefied gas tankers is almost exclusively found in East Asia. In particular, South Korea's large tanker production accounts for 44.2% of global tanker production.

The spatial distribution of shipyard capacity for building large, liquefied gas tankers also raises concerns regarding regional dependencies. To mitigate the geopolitical risk of dependency, other countries could establish large tanker production sites by either converting existing shipyards that are engaged in the building of other vessel types or by initiating the construction of new shipyards. Moreover, shipyards and economies could increase their production to overcome a potential bottleneck in maritime transportation capacity regarding hydrogen Potential solutions to solve the hydrogen transport capacity gap, such as higher shares of local production of hydrogen close to the demand centers, or the long-distance transport of energy carriers via pipelines, are not further examined in this research.

Within this research, various assumptions must be made to project future shipyard capacity requirements and the future hydrogen transport demand. It is therefore assumed that a sufficient supply of auxiliary industries such as tank production will be available. Furthermore, the potential increase in global ammonia demand, whether for use as maritime fuel or to meet rising global fertilizer demand, may be a factor that will influence $LNH_3$ tanker capacity demand and, thus, shipyard capacities in the future. Future research may address this research question. Nevertheless, the research's results underscore the potential bottleneck of maritime hydrogen transport capacity. Consequently, comprehensive analyzes of global energy supply must take these limitations into account. The published results offer the possibility to do so.

.

.

## Declaration of competing interest

The authors declare that they have no known competing financial interest or personal relationships that could have appeared to influence the work reported in this manuscript.

## Acknowledgements

This work was supported by the Helmholtz Association under the program "Energy System Design" and funded by the European Union (ERC, MATERIALIZE, 101076649). Views and opinions expressed are, however, those of the authors only and do not necessarily reflect those of the European Union or the European Research Council Executive Agency. Neither the European Union nor the granting authority can be held responsible for them.

## Author Contribution

Conceptualization: MS, DK, HH; methodology and model development: MS, DK, HH; validation: MS; formal analysis: MS; investigations: MS, DK, HH; writing - original draft: MS, HH; writing – review and editing: MS, DK, HH, JM, JL, GW; visualization: MS, supervision: HH, JL, JM, JL. GW, DS.

All authors have read and agreed to the published version of the manuscript.

Carriers&post_name=Shipyards Shipbuilding&segment_name=a25saQ== (accessed July 17, 2023).

[38] Al-Breiki M, Bicer Y. Comparative cost assessment of sustainable energy carriers produced from natural gas accounting for boil-off gas and social cost of carbon. Energy Reports 2020;6:1897–909.

[39] Brändle G, Schönfisch M, Schulte S. Estimating long-term global supply costs for low-carbon hydrogen. Applied Energy 2021;302:117481. https://doi.org/10.1016/j.apenergy.2021.117481.

[40] IEA. World Energy Outlook 2022. Paris: International Energy Agency; 2022.

[41] IEA. World Energy Outlook 2021. Paris: International Energy Agency; 2021.

[42] Park NK, Suh SC. Tendency toward Mega Containerships and the Constraints of Container Terminals. JMSE 2019;7:131. https://doi.org/10.3390/jmse7050131.

[43] OECD. Long-term baseline projections, No. 109 (Edition 2021). OECD Economic Outlook: Statistics and Projections (Database) 2021. https://doi.org/10.1787/cbdb49e6-en.

[44] Bertagni MB, Socolow RH, Martirez JMP, Carter EA, Greig C, Ju Y, et al. Minimizing the impacts of the ammonia economy on the nitrogen cycle and climate. Proc Natl Acad Sci USA 2023;120:e2311728120. https://doi.org/10.1073/pnas.2311728120.
27

# APPENDIX

## Appendix A: Available shipyard capacities and tanker production capacities of conducted scenarios

A1: Available shipyard capacities and tanker production capacities in *lower hydrogen demand scenario*

| Year | 2030 | 2035 | 2040 | 2045 | 2050 |
|---|---|---|---|---|---|
| Available shipyard capacity $LH_2$ tankers [million CGT] | 11.98 | 9.76 | 16.62 | 19.97 | 10.32 |
| Available shipyard capacity $LNH_3$ tankers [million CGT] | 15.97 | 9.76 | 16.62 | 19.97 | 10.32 |
| Maximum production of $LH_2$ tankers capacity [million $LH_{2eq}$ m³] | 21.68 | 17.66 | 30.06 | 36.13 | 18.67 |
| Maximum production of $LNH_3$ tankers capacity [million $LH_{2eq}$ m³] | 75.92 | 46.38 | 78.97 | 94.91 | 49.04 |
| Maximum number of newbuilt $LNH_2$ tankers | 136 | 111 | 188 | 226 | 117 |
| Maximum number of newbuilt $LNH_3$ tankers | 400 | 245 | 416 | 500 | 259 |

A2: Available shipyard capacities and tanker production capacities in *hydrogen priority scenario*

| Year | 2030 | 2035 | 2040 | 2045 | 2050 |
|---|---|---|---|---|---|
| Available shipyard capacity $LH_2$ tanker [million CGT] | 11.98 | 19.97 | 19.97 | 19.97 | 19.97 |
| Available shipyard capacity $LNH_3$ tanker [million CGT] | 15.97 | 19.97 | 19.97 | 19.97 | 19.97 |
| Maximum production of $LH_2$ tankers capacity [million $LH_{2eq}$ m³] | 21.68 | 36.13 | 36.13 | 36.13 | 36.13 |
| Maximum production of $LNH_3$ tankers capacity [million $LH_{2eq}$ m³] | 75.93 | 94.91 | 94.91 | 94.91 | 94.91 |
| Maximum number of newbuilt $LNH_2$ tankers | 136 | 226 | 226 | 226 | 64 |
| Maximum number of newbuilt $LNH_3$ tankers | 400 | 500 | 500 | 500 | 500 |

A3: Available shipyard capacities and tanker production capacities in *repurpose shipyards scenario*

| Year | 2030 | 2035 | 2040 | 2045 | 2050 |
|---|---|---|---|---|---|
| Available shipyard capacity $LH_2$ tankers [million CGT] | 14.07 | 23.45 | 23.45 | 23.45 | 9.08 |
| Available shipyard capacity $LNH_3$ tankers [million CGT] | 18.76 | 23.45 | 23.45 | 23.45 | 9.08 |
| Maximum production of $LH_2$ tankers capacity [million $LH_{2eq}$ m³] | 21.68 | 36.13 | 36.13 | 36.13 | 10.14 |
| Maximum production of $LNH_3$ tankers capacity [million $LH_{2eq}$ m³] | 75.93 | 94.91 | 94.91 | 94.91 | 26.65 |
| Maximum number of newbuilt $LNH_2$ tankers | 136 | 226 | 226 | 226 | 64 |
| Maximum number of newbuilt $LNH_3$ tankers | 400 | 500 | 500 | 500 | 141 |

A4: Available shipyard capacities and tanker production capacities in *crude oil inclusion scenario*

| Year | 2030 | 2035 | 2040 | 2045 | 2050 |
|---|---|---|---|---|---|
| Available shipyard capacity $LH_2$ tankers [million CGT] | 18.34 | 30.57 | 28.82 | 28.82 | 14.46 |
| Available shipyard capacity $LNH_3$ tankers [million CGT] | 24.45 | 30.57 | 28.82 | 28.82 | 14.46 |
| Maximum production of $LH_2$ tankers capacity [million $LH_{2eq}$ m³] | 33.18 | 55.31 | 52.15 | 52.15 | 26.16 |
| Maximum production of $LNH_3$ tankers capacity [million $LH_{2eq}$ m³] | 116.23 | 145.29 | 136.98 | 136.98 | 68.72 |
| Maximum number of newbuilt $LNH_2$ tankers | 208 | 346 | 326 | 326 | 164 |
| Maximum number of newbuilt $LNH_3$ tankers | 612 | 765 | 722 | 722 | 362 |

A5: Available shipyard capacities and tanker production capacities in *container inclusion scenario*

| Year | 2030 | 2035 | 2040 | 2045 | 2050 |
|---|---|---|---|---|---|
| Available shipyard capacity $LH_2$ tankers [million CGT] | 3.78 | 6.31 | 6.31 | 6.31 | -8.05 |
| Available shipyard capacity $LNH_3$ tankers [million CGT] | 5.05 | 6.31 | 6.31 | 6.31 | -8.05 |
| Maximum production of $LH_2$ tankers capacity [million $LH_{2eq}$ m³] | 6.85 | 11.42 | 11.42 | 11.42 | 0 |
| Maximum production of $LNH_3$ tankers capacity [million $LH_{2eq}$ m³] | 24.00 | 30.01 | 30.01 | 30.01 | 0 |
| Maximum number of newbuilt $LNH_2$ tankers | 43 | 72 | 72 | 72 | 72 |
| Maximum number of newbuilt $LNH_3$ tankers | 127 | 158 | 158 | 158 | 0 |

# Appendix B: Future maritime hydrogen transport capacity demand

B1: Scenario dependent derived maritime hydrogen transport capacity demand for between 2030 and 2050 in million m³ LH$_{2eq}$ cargo capacity

| Year | 2030 | 2035 | 2040 | 2045 | 2050 |
|---|---|---|---|---|---|
| *LNG-first scenario* | 51.69 | 71.39 | 91.09 | 110.78 | 130.48 |
| *Lower hydrogen demand scenario* | 32.00 | 39.39 | 46.78 | 54.16 | 61.55 |
| *Hydrogen priority scenario* | 51.69 | 71.39 | 91.09 | 110.78 | 130.48 |
| *Repurpose shipyards scenario* | 51.69 | 71.39 | 91.09 | 110.78 | 130.48 |
| *Crude oil inclusion scenario* | 51.69 | 71.39 | 91.09 | 110.78 | 130.48 |
| *Container inclusion scenario* | 51.69 | 71.39 | 91.09 | 110.78 | 130.48 |